\documentclass[a4paper,10pt]{article}
\usepackage{epsfig,amsmath,amssymb,verbatim,mathrsfs,array,layout,textcomp,amssymb,latexsym}
\usepackage{subfigure}
\usepackage{epstopdf}
\usepackage{cite}
\newenvironment{Eqnarray}{\arraycolsep 0.14em\begin{eqnarray}}{\end{eqnarray}}
\def\beqa{\begin{Eqnarray}}
\def\eeqa{\end{Eqnarray}}
\def\beq{\begin{Eqnarray}}
\def\eeq{\end{Eqnarray}}

\newcommand{\bv}{\left(\begin{array}{c}}
\newcommand{\ev}{\end{array}\right)}
\newcommand{\bmtwo}{\left(\begin{array}{cc}}
\newcommand{\bmthree}{\left(\begin{array}{ccc}}
\newcommand{\emn}{\end{array}\right)}
\newcommand{\bmtwoc}{\left\{\begin{array}{cc}}
\newcommand{\bmthreec}{\left\{\begin{array}{ccc}}
\newcommand{\emnc}{\end{array}\right\}}

\usepackage{amsmath}
\usepackage{amssymb}
\usepackage{slashed}
\usepackage{array}
\newcommand{\hmass}{\ensuremath{\mathrm{125~GeV}}}

\newcommand{\pt}{\ensuremath{p_T}}
\newcommand{\hte}{\ensuremath{h \rightarrow \tau^{\pm} e^{\mp}}}
\newcommand{\htm}{\ensuremath{h \rightarrow \tau^{\pm} \mu^{\mp}}}

\newcommand{\ztt}{\ensuremath{Z \rightarrow \tau^{+} \tau^{-} }}
\newcommand{\ttbar}{\ensuremath{t \bar{t}}}
\newcommand{\m}{\ensuremath{\mu}}
\newcommand{\tov}{\ensuremath{\tau}}
\newcommand{\mcol}{\ensuremath{m_{\rm coll}}}
\newcommand{\lumi}{\ensuremath{\mathrm{20~fb^{-1}}}}
\newcommand{\cme}{\ensuremath{\mathrm{8~TeV}}}
\newcommand{\mes}{\ensuremath{\mu e}}
\newcommand{\ems}{\ensuremath{e\mu}}

\begin{document}

\vskip1.5cm
\begin{center}
{\Large \bf Asymmetric lepton-flavor violating Higgs decays}
\end{center}
\vskip0.2cm

\begin{center}
Shikma Bressler, Avital Dery and Aielet Efrati \\
\end{center}
\vskip 8pt
\begin{center}
{\it Department of Particle Physics and Astrophysics \\
Weizmann Institute of Science, Rehovot 7610001, Israel} \vspace*{0.3cm}
\vskip 8pt
{\tt shikma.bressler,avital.dery,aielet.efrati@weizmann.ac.il}
\end{center}
\vskip 8pt


\begin{abstract}
We introduce a new method to search for the lepton-flavor violating Higgs decays \htm~and \hte~in the leptonic $\tau$~decay channel. In particular, the Standard Model background is estimated in a fully data driven way. The method exploits the asymmetry between electrons and muons in the final state of signal events and is sensitive to differences in the rates of the two decays. Using this method, we investigate the LHC sensitivity to these processes. With \lumi~of data at $\sqrt{s}=8$~TeV, we expect a $3\sigma$ sensitivity for observing branching ratios of order $0.9\%$. The method and the suggested statistical treatment are discussed in detail.
\end{abstract}

\vspace{1cm}

\section{Introduction}\label{sec:introduction}

Recent results of both the ATLAS~\cite{HiggsATLAS} and CMS~\cite{Chatrchyan:2013mxa,Chatrchyan:2013iaa,Chatrchyan:2014vua} collaborations support the assumption that the new resonance discovered at the vicinity of \hmass~is indeed a Higgs boson related to the mechanism of electroweak symmetry breaking, with couplings that are not very different from the Standard Model (SM) predictions. These studies focus on precise measurements of attributes predicted by the SM; primarily, couplings to gauge bosons and same flavor fermions. A complementary and well motivated approach is the search for non-SM properties of this \hmass~particle. Among these, Lepton-Flavor Violating (LFV) couplings form an interesting class~\cite{Blankenburg:2012ex,Harnik:2012pb,Davidson:2012ds,Celis:2013xja,Dery:2013rta}.

LFV is known to exist in nature. The observation of neutrino oscillations indicates that lepton flavor is not an exact symmetry, and calls for physics beyond the SM that participates in flavor changing dynamics. With the recently discovered Higgs boson, new channels in which such dynamics may be observed become experimentally available. LFV Higgs decays are expected in many extensions of the SM~\cite{Han:2000jz,Cotti:2001fm,Kanemura:2005hr,GomezBock:2005hc,Giudice:2008uua,Davidson:2010xv,Botella:2011ne,Cely:2012bz,Arhrib:2012ax,McKeen:2012av,Giang:2012vs,Arana-Catania:2013xma,Falkowski:2013jya,Adachi:2014wva}; The general supersymmetric SM Lagrangian, for instance, includes various LFV terms~\cite{Assamagan:2002kf,Brignole:2003iv,Brignole:2004ah,Arcelli:2004af}. The Froggatt-Nielsen mechanism and the general two-Higgs doublet models are other examples in which Higgs LFV decay modes arise~\cite{Froggatt:1978nt,Dery:2013rta,Branco:2011iw,Paradisi:2005tk}. In this work we study such Higgs decays.

We consider the following mass basis Lagrangian for the Yukawa interactions:
\beqa
-\mathcal{L}_Y=\frac{c_{ij}}{\sqrt{2}}h\bar\ell^i_L\ell^j_R+{\rm h.c.}\,,\;\;\;i,j=e,\mu,\tau\,,
\eeqa
where $c_{ij}^{\rm SM}=\delta_{ij}\sqrt{2}m_i/v$ and $v=246$~GeV. The current bounds on the LFV couplings $c_{ij}$ (with $i\neq j$) are indirect, the strongest of which are derived from the upper limits on the processes $\mu \rightarrow e \gamma$~\cite{Adam:2013mnn}, $\tau \rightarrow \mu \gamma$ and $\tau \rightarrow e \gamma$. (See \cite{Blankenburg:2012ex,Harnik:2012pb}, and references within.) In the absence of cancelations, these bounds constrain $|c_{\mu e}|$ (and $|c_{e\mu}|$) to be very small, such that the decay $h \rightarrow \mu^{\pm}e^{\mp}$ is not likely to be observed at the LHC. On the other hand, the inferred bounds on $|c_{\tau \mu}|$ and $|c_{\tau e}|$ ($|c_{\mu\tau}|$ and $|c_{e\tau}|$) are much weaker, allowing for ${\rm BR}\left(h\to\tau^\pm\mu^\mp\right)$ or ${\rm BR}\left(h\to\tau^\pm e^\mp\right)$ as high as ${\cal O}(10\%)$. The upper limit on the rate of $\mu \rightarrow e \gamma$ imposes an additional strong bound on the product $|c_{\tau\mu}c_{e\tau}|$ (and $|c_{\tau\mu}c_{e\tau}|$) suggesting that if one coupling is significantly larger than zero, the other is essentially zero. Thus, the two processes, \htm~and \hte~, are weakly bounded separately but are not expected to coexist in observable rates. Limits on tree-level processes, such as $\tau\rightarrow3\mu$, result in much weaker bounds.

The LHC sensitivity to the decays \hte~and \htm~was studied independently in Ref.~\cite{Harnik:2012pb} for hadronic $\tau$ decays and in Ref.~\cite{Davidson:2012ds} for leptonic $\tau$ decays. Upon application of selection criteria to enhance the signal to background ratio, both studies predict that a ${\rm BR}$ of order of $0.45\%$ could be excluded at 95\% C.L.. Both approaches implicitly rely on background estimation from Monte-Carlo simulation but do not attempt to address the issue of the large systematic uncertainties it may entail. Therefore, from the experimental point of view, the main remaining challenge lies in obtaining a good estimation for the SM background.

In this work we address the task of modeling the background for the leptonic $\tau$ decay selection. We present a fully data driven method that makes use of two mutually exclusive data samples; each sample serves, in a way, as a background estimation for the other. Our approach is sensitive to an asymmetry between the two rates ${\rm BR}\left(h\to\tau^\pm\mu^\mp\right)$~and ${\rm BR}\left(h\to\tau^\pm e^\mp\right)$, henceforth denoted as ${\rm BR}_{\tau\mu}$ and ${\rm BR}_{\rm \tau e}$. It can further be easily adjusted for the searches of LFV decays of other resonances. The method evades the large systematic uncertainties expected in traditional background estimation techniques for these channels. The leading uncertainty is expected to be governed by statistics, thus the obtained sensitivity will improve with more LHC data.

This paper is organized as follows: our background estimation method is described in section~\ref{sec:theMethod}, followed by a detailed description of the statistical procedure in section~\ref{sec:statistics}. In section~\ref{sec:sensitivity} we present the expected sensitivity of the LHC to LFV Higgs decays. We conclude the work in section~\ref{sec:summary}.

\section{The background estimation method}\label{sec:theMethod}

In order to convey the need for an unorthodox background estimation method, we first sketch the event selection and list the dominant background processes and their relation to the signal distributions. We define the signal processes as Higgs decays to $\tau^\pm\mu^\mp$ and to $\tau^\pm e^\mp$, where the $\tau^\pm$ further decays to the other flavor lepton. The resulting final state consists of one electron and one muon of opposite sign, and missing transverse energy $(\slashed E_T)$. At leading order, it is also characterized by low jet activity.  Consequently, the leading SM background process for this selection is $Z\rightarrow \tau^+\tau^-\rightarrow\mu^\pm e^\mp+\slashed E_T$, with kinematics closely resembling those of the signal. The full list of background processes, along with production cross-sections and the expected numbers of final $e^\pm\mu^\mp$ events before further cuts, are listed in Table~\ref{tab:mcSamples}.

\begin{figure}
  \centering
   \includegraphics[width=2.7in]{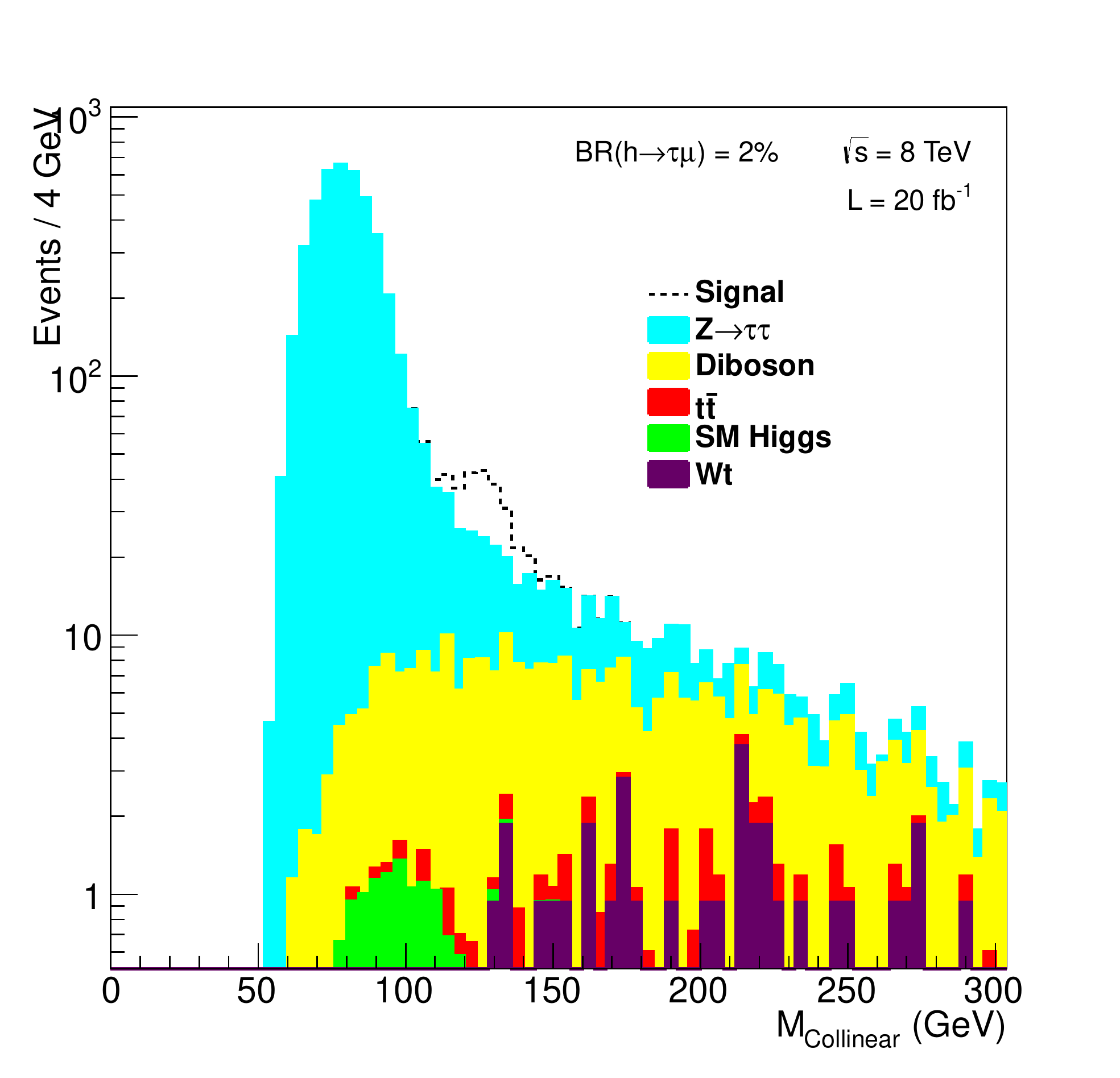}
  \caption{The Collinear mass distribution of the various SM background processes and a signal of ${\rm BR}_{\tau\mu}=2\%$ after selection cuts (Monte-Carlo simulation, for detailed description see Table~\ref{tab:cutFlow}). The contributions of the various processes are stacked on top of each other.}
  \label{fig:mcol}
\end{figure}

The reconstructed mass distributions, $m_{\rm coll}$, for the simulated signal and background samples, after selection cuts following Ref.~\cite{Davidson:2012ds} (detailed in Table~\ref{tab:cutFlow}) is shown in Figure~\ref{fig:mcol}. Evidently, the signal lies in a transitional region between i) the dominant $Z$ peak and ii) the di-boson contribution that becomes prominent at higher mass values. Background modeling techniques based on extrapolation from regions outside the Higgs mass window are therefore not well suited for this channel. Monte-Carlo based estimation is also likely to suffer from large uncertainties arising from the poor modeling of the tail of the $Z$  mass distribution. Moreover, the close resemblance of the signal and the \ztt~kinematics would make its validation difficult.
\begin{table}[b!]
  \begin{center}  	
  \small
    \begin{tabular}{| l | c | c |}
\hline
Process & cross-section (pb) & number of $e^\pm\mu^\mp$ events \\
\hline
\hline
\hte~(with ${\rm BR}_{\tau e}=1\%$) & 0.2~\cite{Heinemeyer:2013tqa} & 186 \\
\htm~(with ${\rm BR}_{\tau\mu}=1\%$) & 0.2~\cite{Heinemeyer:2013tqa} & 186 \\
\hline
$Z/\gamma^*\rightarrow\tau\tau$ ($M_{\ell\ell}>60$~GeV) &1111.3~\cite{Gavin:2010az,Gavin:2012sy} & 160801  \\
\ttbar & 226.6~\cite{Cacciari:2011hy,Czakon:2012zr,Czakon:2012pz} & 42097 \\
$WW$ & 57.2~\cite{Campbell:2011bn} & 11710 \\
$Z/\gamma^* Z/\gamma^*$ & 7.9~\cite{Campbell:2011bn} & 132 \\
$WZ/\gamma^*$ & 22.9~\cite{Campbell:2011bn} & 573 \\
SM $h$ & 19.3~\cite{Heinemeyer:2013tqa} & 896\\
$Wt$ & 22.2~\cite{Kidonakis:2013zqa} & 7474\\
\hline
    \end{tabular}
  \end{center}
  \caption{The signal and background simulated samples, before cuts. The production cross sections are for a center of mass energy of \cme, taken from the latest available theoretical estimations. These are used to normalize the number of events to match $\mathcal{L}=20~{\rm fb}^{-1}$. The samples were generated using Pythia8~\cite{Sjostrand:2007gs} (with MSTW2008 pdf set) and the Delphes3 software~\cite{deFavereau:2013fsa} with the standard ATLAS card configuration. The $Wt$ sample was generated using MadGraph 5~\cite{Alwall:2011uj}. We assume SM Higgs production and total width and use $m_t=173.3$~GeV.}
  \label{tab:mcSamples}
\end{table}
\begin{table}[h!]  \begin{center}
\small
  	\begin{tabular}{| p{4cm} || p{1cm} || p{1.5cm} | p{1.2cm} | p{0.8cm} | p{0.8cm} | p{0.8cm} || p{1cm} |}
\hline
Cut & $N_{\text{bg.}}$ & $N_{Z/\gamma^*\to\tau\tau}$ & $N_{\text{Diboson}}$ & $N_{t\bar{t}}$ & $N_{Wt}$ & $N_{h^{\text{SM}}}$ & $N_{\text{signal}}$  \\
\hline
\hline
Exactly 2 opp. sign diff. flavor leptons (e \m) & 173733 & 109666 & 12415 & 43282 & 7474 & 896 & 186\\
\hline
$ p_T^{\ell_1} > 20 \quad\rm{GeV}$ & 80004 & 29931 & 9607 & 34342 & 5727 & 398 & 123\\
$ p_T^{\ell_0} > 30 \quad\rm{GeV}$ & 66489 & 18229 & 9050 & 33329 & 5554 & 326 & 121\\
\hline
jet veto: no jet with $p_T > 30 \quad\rm{GeV}$ and $|\eta|<2.5$ & 22340 & 14000 & 6587 & 736 & 914 & 170 & 71\\
\hline
$\Delta\phi(\ell_0,\ell_1)>2.5$ & 17083 & 13240 & 3107 & 395 & 280 & 61 & 64\\
$\Delta\phi(\ell_1,\slashed{E}_T)<0.5$ & 6864 & 6002 & 657 & 96 & 83 & 25 & 50\\
\hline
Efficiency & 3.9\% & 5.5\% & 5.3\% & 0.2\% & 1.1\% & 10.1\% & 26.7\%\\
\hline
    \end{tabular}
  \end{center}
  \caption{Selection cuts, following Ref.~\cite{Davidson:2012ds}, and remaining number of signal and background events after each cut.}
  \label{tab:cutFlow}
\end{table}

We introduce an alternative method, relying solely on data; hence, this approach is not subject to any theoretical or simulation mismodeling uncertainties. For simplicity we first present the concept under the assumption that ${\rm BR}_{\tau e}=0$ and ${\rm BR}_{\tau\mu}\neq0$ and later generalize for scenarios where both rates are non zero. Our method is based on the following two premises:
\begin{enumerate}
\item SM processes at the LHC center-of-mass energy result in data that is approximately symmetric to the replacement of prompt electrons with prompt muons. That is, the kinematic distributions of  prompt electrons and prompt muons are approximately the same.
\item \htm~decays break this symmetry.
\end{enumerate}
The second premise can be justified rather easily; since the muons are produced in the Higgs decay and the electrons are produced in the decay of the \tov, the \pt~of the electrons is typically softer than that of the muons.

We define the \textbf{leading lepton}, $\ell_0$, and the \textbf{sub-leading lepton}, $\ell_1$, to be the leptons with the higher and lower \pt~respectively, and write an expression for the collinear mass in terms of $p^{\ell_0}_{T}$~and $p^{\ell_1}_{T}$:
\begin{equation}
m_{\rm coll} = \sqrt{2p_T^{\ell_0}\left(p_T^{\ell_1}+\slashed{E}_T\right)\left(\cosh\Delta\eta-\cos\Delta\phi\right)}\,,
\end{equation}
where $\Delta\eta,\,\Delta\phi$ are the corresponding angles between the leading and sub-leading leptons. We divide the data into two mutually exclusive samples:
\begin{itemize}
\item $\boldsymbol\mu\mathbf{e}$ \textbf{data sample}: $\ell_0$~is the muon and $\ell_1$~is the electron ($p_T^\mu \geq p_T^e$)
\item $\mathbf{e}\boldsymbol\mu$ \textbf{data sample}: $\ell_0$~is the electron and $\ell_1$~is the muon ($p_T^e > p_T^\mu$)
\end{itemize}
According to our premises, the SM background is divided equally between the two samples. However, since the electron \pt~spectrum of the signal is typically softer than the muon \pt~spectrum, the vast majority of the \htm~events are expected to be found in the \mes~sample.

\begin{figure}
  \centering
   \includegraphics[width=2.7in]{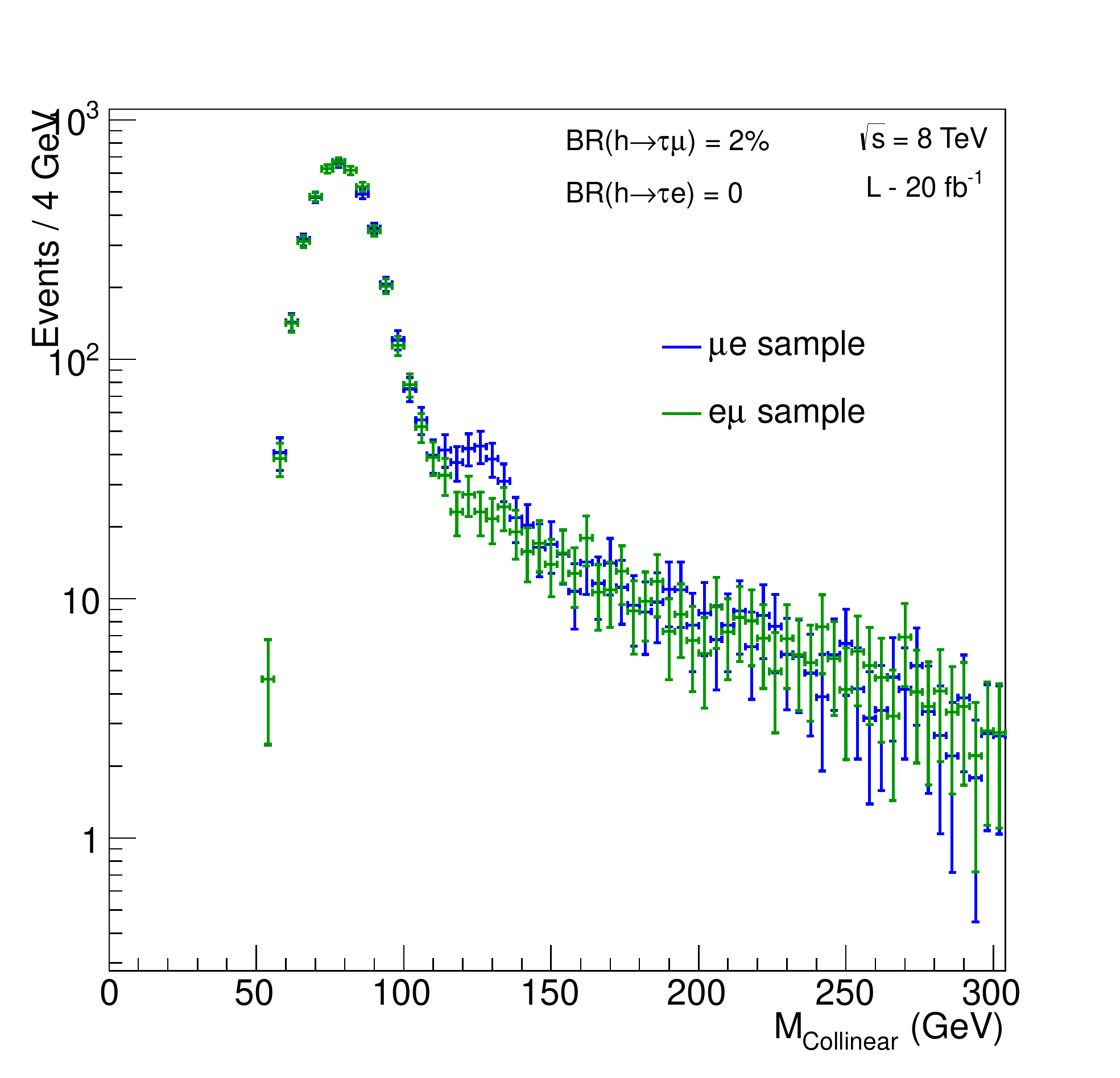}
  \caption{Collinear mass distributions of the $\mu e$ (blue) and $e\mu$ (green) data samples with ${\rm BR}_{\tau\mu}=2\%$ and ${\rm BR}_{\tau e}=0$ (Monte-Carlo simulation).}
  \label{fig:mcol_EMME}
\end{figure}

The result is two data samples for which the SM kinematic variables, in terms of $\ell_0$ and $\ell_1$, are similarly distributed. In particular, the collinear mass distribution in the two data samples is approximately the same for SM processes. However, signal events are distributed asymmetrically, appearing only in the \mes~sample as a relatively narrow peak in the collinear mass distribution, on top of the SM background distribution which is now, to a good approximation, well modeled by the $m_{\rm coll}$ distribution of the \ems~sample. This can be seen in Figure~\ref{fig:mcol_EMME}, in which the SM background processes and a signal of ${\rm BR}_{\tau\mu}=2\%$ were used to generate \mes~and \ems~collinear mass distributions, appearing on top of each other. The signal region, between 100 and 150 GeV, exhibits a clear asymmetry between the two samples, supporting our second premise. Outside of the signal region, the \mes~distribution is well modeled by the \ems~distribution, supporting our first premise.

While the second premise requires very little explanation, justifying the first one is more demanding. The SM gauge interactions are universal, meaning that coupling strengths of gauge bosons to electrons and muons are identical. Variation in the rates of gauge processes involving different leptons is only due to phase space effects. Yukawa interactions of the SM, on the other hand, exhibit non-universality even prior to phase space effects, since the Yukawa couplings are proportional to the lepton masses. Nevertheless, at energy scales relevant for the LHC, both Yukawa and phase space induced differences affect the symmetry very little. This is the basis of our symmetry argument. Moreover, since our procedure requires the selection of one electron and one muon in the final state, the effects of these small sources of asymmetry on the collinear mass distribution cancel out.

From an experimental point of view, electrons and muons are different objects. Once again, the requirement of both an electron and a muon in the final state ensures the robustness of the symmetry argument; most potential sources of asymmetry will have a similar effect on the \mes~and \ems~samples. One factor that could potentially degrade the electron-muon symmetry is the difference in reconstruction and trigger efficiency values. We argue that as long as the event selection requires electrons and muons with \pt~above the efficiency plateaus, these differences do not induce asymmetry between the two data samples; the effect on the event reconstruction efficiency in both samples will be a multiplication of two efficiency values. A selection that includes lower \pt~electrons and muons can be corrected to restore the symmetry by introducing additional scale factors. Another potential source of asymmetry is the energy loss of electrons in Bremsstrahlung radiation, which softens their \pt~spectrum compared to that of muons. This effect may introduce differences between the two data samples. Yet, we expect it to generate only a small asymmetry, since it affects the two reconstructed mass distributions in a similar way.

Fake and non-prompt leptons (from W+jets and QCD events) are expected to contribute differently to the two samples since the origin of fake and non prompt muons is different than that of electrons. Non prompt muons, for instance, can originate from $\pi^{\pm}$~decays while fake electron can arise from photon conversion. The effect of fake and non-prompt leptons is not modeled in our simulation. Both ATLAS and CMS have developed methods to accurately estimate contamination of non-prompt leptons in various signal regions~\cite{Aad:2010ey,Chatrchyan:2013fea,Aad:2012mea}. These methods can be used to estimate the number of fake events in the \mes~and \ems~samples.

An important advantage of this background estimation method is the large number of control regions that can be used to study the symmetry assumption and control the systematic uncertainties associated with it. The high ($\geq 150$~GeV) and low ($\leq 100$~GeV) mass regions are natural control regions. Selection criteria could be reversed to generate additional control regions.

While we have so far only discussed a signal in the \htm~channel, our approach is not limited to the scenario where $\Gamma(h\rightarrow \tau e) = 0$. The background estimation method we present enables sensitivity to the difference between the two decay rates. We note, however, that the unfavorable case in which $\Gamma(h\rightarrow \tau\mu) \simeq \Gamma(h\rightarrow \tau e)$ does not generate an asymmetry between the two samples and therefore requires a different approach.

\section{Statistical treatment}\label{sec:statistics}

As argued above, in the absence of asymmetric LFV Higgs decays, the collinear mass distribution of the \ems~and \mes~data sets originates from the same distribution, denoted as $B$. The differences between the two distributions are then only due to statistical fluctuations. Therefore, we estimate the number of background events in each bin $i$ ($B_i$) by maximizing the likelihood function with respect to the different $B_i$'s and signal strength $\mu$
\begin{equation}
L(B_i,\mu)=\prod_i {\rm Pois}(n_i\mid B_i)\times {\rm Pois}(m_i\mid B_i+\mu s_i)\,.
\end{equation}
Here $n_i$, $m_i$ and $s_i$ are the number of \ems~events, \mes~events and signal events in bin $i$, respectively. We model a Gaussian shaped signal around $m=125.8$~GeV with $\Gamma\simeq8.6$~GeV, where the mass and width are fitted to the simulated signal distribution. A unit strength $\mu=1$ is compatible with ${\rm BR}_{\tau\mu}= 1\%$. If both decay rates are non zero, $\mu$ corresponds to the difference between the two branching ratios. Since statistical fluctuations are Poisson distributed, the error on each $B_i$ is estimated as $\sigma_{B_i} = \frac{1}{2}\sqrt{n_i+m_i}$. For simplicity we do not assign systematic uncertainty to the symmetry assumption. However, the model can be generalized to include such uncertainties.

\begin{figure}[b!]
  \centering
   \includegraphics[width=2.7in]{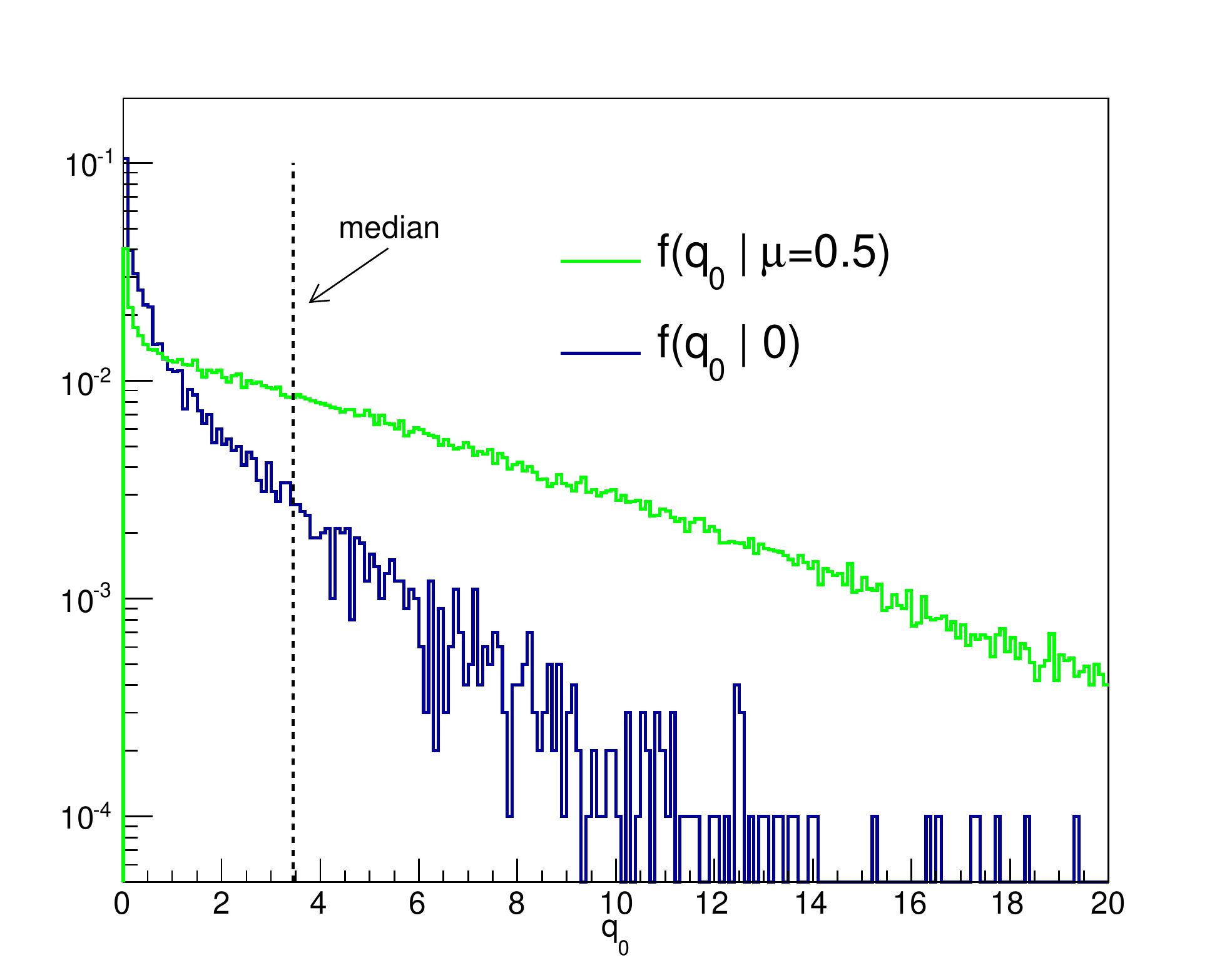}
  \caption{The distributions $f(q_0\mid 0)$ (blue) and $f(q_0\mid \mu=0.5 )$ (green). The $p$-value for the rejection of the null hypothesis is given by the area under the curve $f(q_0\mid 0)$ to the right of the median of $f(q_0\mid \mu)$.}
  \label{fig:pdfs}
\end{figure}

We estimate the LHC sensitivity using the CLs method with profile likelihood ratio as a test statistic~\cite{Cowan:2010js}. Looking at the collinear mass distribution we construct the likelihood function of our problem as follows
\begin{equation}
L(\mu, b_i) = \prod_i {\rm Pois}(n_i\mid b_i)\times {\rm Pois}(m_i \mid b_i + \mu s_i)\times {\rm Gauss}(b_i \mid B_i , \sigma_{B_i})
\end{equation}
With $n_i$, $m_i$, $B_i$, $\sigma_{B_i}$, $s_i$ and $\mu$ as defined in the previous paragraph. $b_i$ is the background in each bin and it is treated as nuisance parameter, Gaussian distributed around $B_i$ with variance $\sigma_{B_i}$. We use $q_\mu=-2\ln\left[L\left(\mu,\hat{\hat\theta}\right)/L\left(\hat\mu,\hat\theta\right)\right]$ and extract the pdf's $f(q_{\mu}\mid\mu^\prime)$ (under the assumption of the signal strength $\mu^\prime$) from large sets of toy Monte-Carlo samples. For illustration, Figure~\ref{fig:pdfs} shows the two pdf's $f(q_{0}\mid 0)$~and $f(q_{0}\mid \mu)$ for ${\rm BR}_{\tau\mu}= 0.5\%$. The $3\sigma$ sensitivity for discovery is determined as the value of $\mu$ for which the median of $f(q_{0}\mid \mu)$ satisfies $\int^{\infty}_{\rm q_{\rm med}}f(q_{0}\mid 0) {\rm d}q_0= 2.7\times10^{-3}$.

\section{Results: the expected sensitivity}
\label{sec:sensitivity}

To estimate the expected sensitivity at the LHC we simulated the signal and SM background processes using Pythia8~\cite{Sjostrand:2007gs} (with MSTW2008 pdf set). The $Wt$ sample was generated using MadGraph 5~\cite{Alwall:2011uj}. These processes, along with the corresponding production cross-sections and the expected numbers of final $e^\pm\mu^\mp$ events are listed in Table~\ref{tab:mcSamples}. For the production cross sections we used the latest available theoretical estimations (the relevant Refs. are also listed in Table~\ref{tab:mcSamples}). Detector effects were modeled using the Delphes3 software~\cite{deFavereau:2013fsa} with the standard ATLAS card configuration. In all the samples we assume SM Higgs production and total width and use $m_t=173.3$~GeV. The selection criteria and the number of remaining background and signal events after each step are detailed in Table~\ref{tab:cutFlow}. The distribution of \mcol~after applying the entire cut-flow is shown in Figure~\ref{fig:mcol}, where an \htm~signal assuming a 2\% branching ratio is shown in black on top of the various SM processes.

\begin{figure}[h!]\label{fig:sensitivity}
  \begin{center}
      \subfigure[]{\scalebox{1.4}{\includegraphics[width=2.7in]{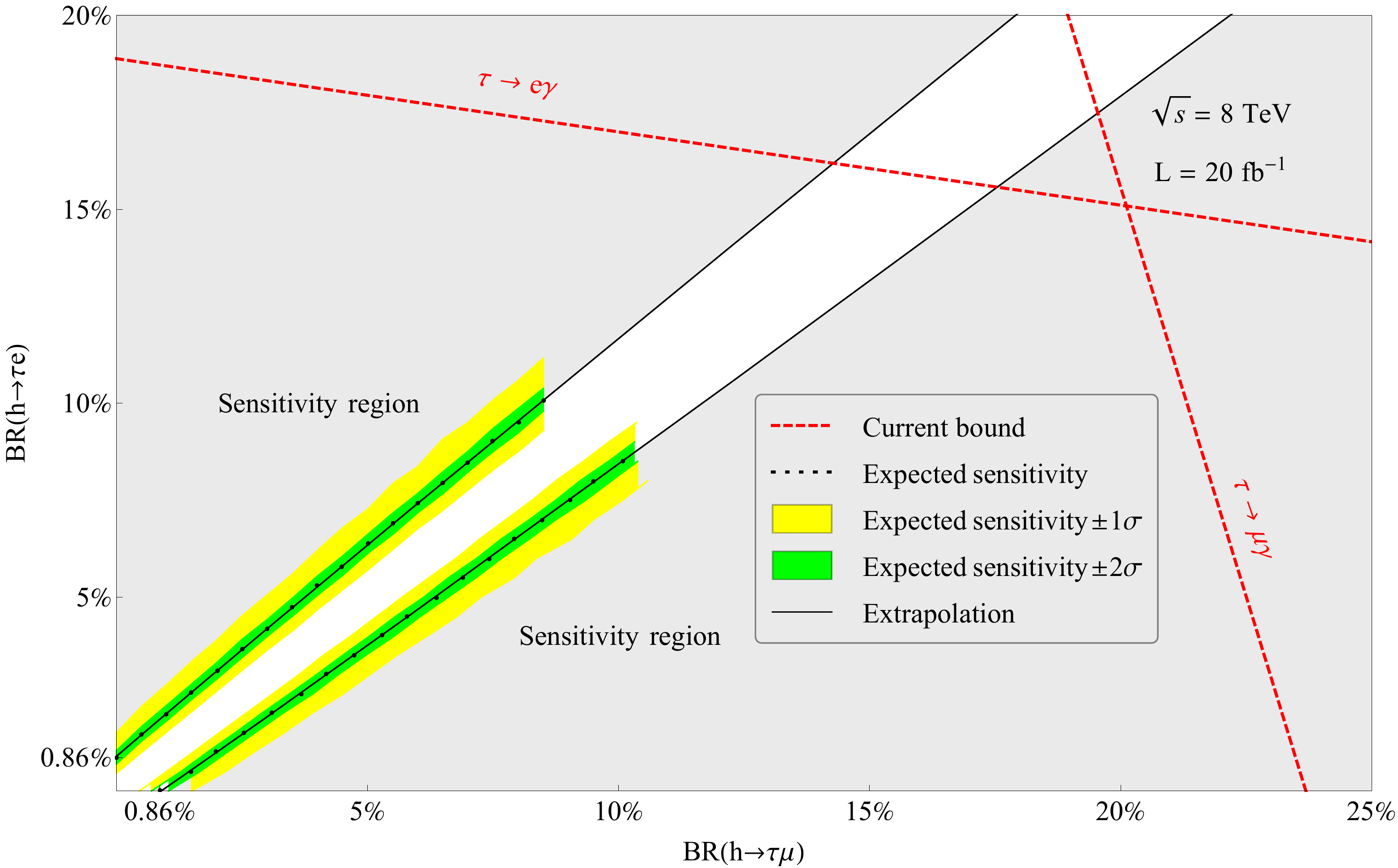}}\label{fig:a}}
      \subfigure[]{\scalebox{1.4}{\includegraphics[width=1.7in]{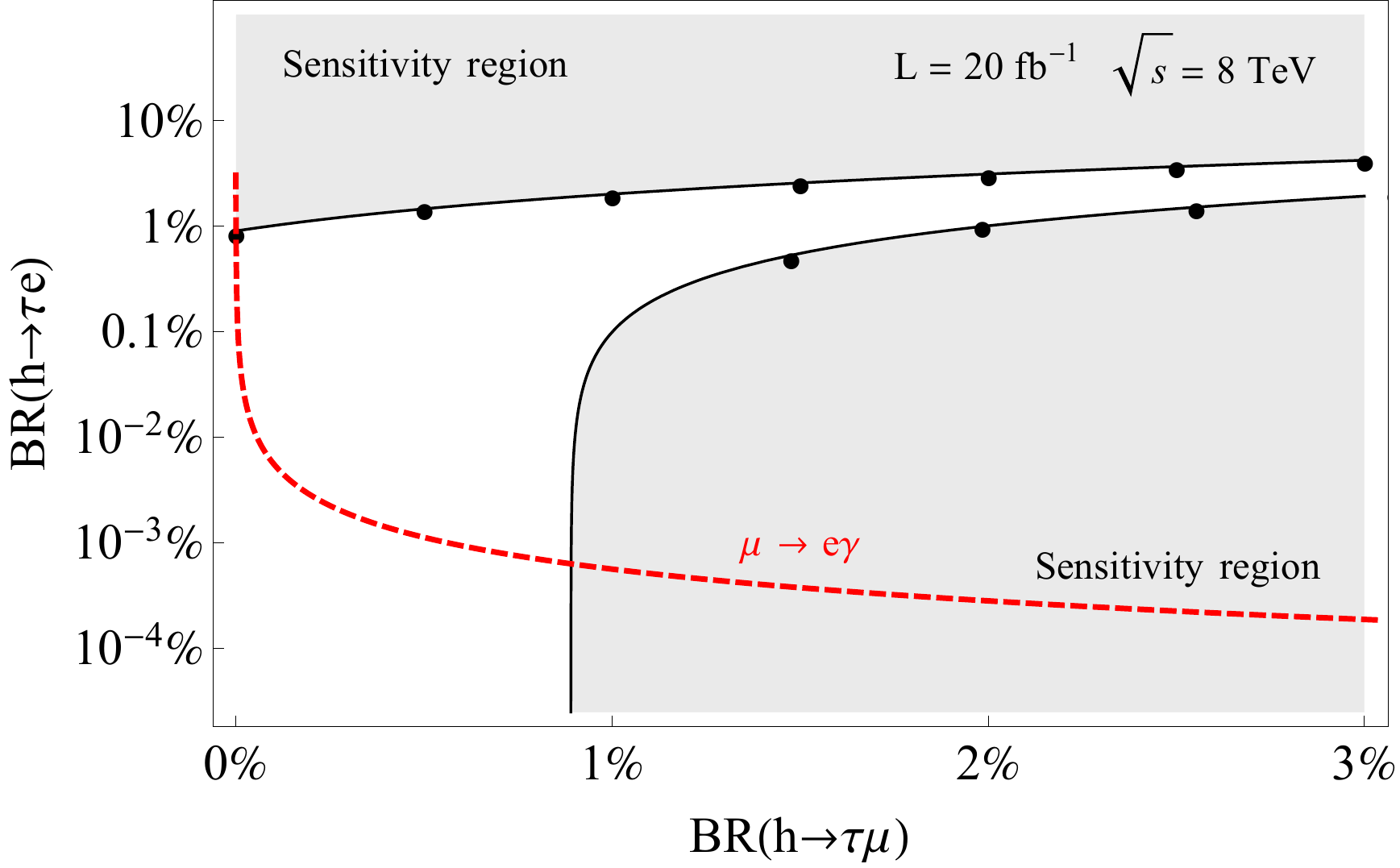}}\label{fig:b}}
      \caption{The expected $3\sigma$ sensitivity for discovery of  LFV Higgs decays at the LHC with $\sqrt{s}=8$~TeV and $\mathcal{L}=20~{\rm fb}^{-1}$. black dots stand for the numerical results for the expected median sensitivity, with $1\sigma$ (green) and $2\sigma$ (yellow) bands. The solid lines correspond to the extrapolation (using a two parameter fit to the numerical results). The dashed red lines show the current indirect bounds~\cite{Adam:2013mnn,Beringer:1900zz}.}
  \end{center}
\end{figure}

We present our results in Figure~\ref{fig:sensitivity}, which shows the expected $3\sigma$~sensitivity for discovering \htm~and \hte~decays in \lumi~of data at $\sqrt{s}=8$~TeV. It is symmetric with respect to ${\rm BR}_{\tau\mu}$ and ${\rm BR}_{\tau e}$, as the method probes the asymmetry between these two decay modes. The median values for the expected sensitivity are shown in black dots,
defining a large sensitivity region (in the BR plane) between these points and the axes; the green and yellow bands show the $1\sigma$ and $2\sigma$ ranges around the expected median
values, corresponding to statistical fluctuations of the data. Note the lack of sensitivity to the scenario of equal branching ratios. This is an inherent feature of the method, as previously discussed.

We find that if ${\rm BR}_{\tau e}=0$ but ${\rm BR}_{\tau\mu}\neq0$ is realized in nature, an \htm~branching ratio of $0.86\%$ can be observed at the $3\sigma$ level with the current data collected by
the LHC. If both ${\rm BR}_{\tau\mu}$ and ${\rm BR}_{\tau e}$ exist, the sensitivity to the asymmetry between the resulting decay modes is slightly lower because the same number of signal (asymmetric)
events have to be seen on top of a slightly larger (symmetric) background. Qualitatively, the expected sensitivity would then be given by a constant contour of
$\left({\rm BR}_{\tau\mu}-{\rm BR}_{\tau e}\right)s/\sqrt{B+{\rm BR}_{\tau\mu}s}$. This expected behavior is found to be in excellent agreement with the numerical results, as can be seen in Figure~\ref{fig:sensitivity} (where a two parameter numerical fit to the data is implemented). Any observation of a signal will be degenerate along such line. However, we emphasize that the scenario where both decay rates are non negligible is extremely unfavorable due to the stringent bound arising from $\mu\rightarrow e\gamma$ searches~\cite{Adam:2013mnn}. (See Figure~\ref{fig:b}.)

We tested the effect of worse $\slashed E_T$ resolution on the sensitivity. When widening its distribution by 20\%, on top of the Delphes simulation, the expected $3\sigma$ sensitivity for the scenario ${\rm BR}_{\tau e}=0$ and ${\rm BR}_{\tau\mu}\neq0$ degrades to 0.97\%.

Since the background estimation method is fully data driven, the leading systematic uncertainty is a result of the limited statistics. This effect will decrease with more LHC data; the statistical nature of the method dictates that its sensitivity will improve as $1/\sqrt{L}$. We verified this behavior numerically. Various smoothing techniques could also be used to further reduce this uncertainty.

\section{Summary}\label{sec:summary}

Lepton-flavor violating processes are predicted by many extensions of the SM. The observation of neutrino oscillations proves their existence and motivates the search for other such processes. With LHC data continuing to accumulate, Higgs LFV decays may become experimentally available.

In this work we develop a fully data driven method to estimate the SM background in the search for the LFV higgs decays \htm~and \hte. It relies on the asymmetry between electrons and muons in the final state of signal events and not on specific properties of the Higgs particle, which are commonly used to enhance the signal to background ratio but are not essential to the procedure. Therefore, it can be generalized for analogous LFV decay rates of any other known or hypothetical particle such as the $Z$~boson, additional $Z^\prime$ or any doubly-charged resonance.

The method is sensitive to differences between the decay rates $\Gamma(h\rightarrow \tau\mu)$~and $\Gamma(h\rightarrow \tau e)$. Under the assumption that one of the decay rates is negligibly small, we predict a $3\sigma$~sensitivity for discovering ${\rm BR}_{\tau\mu}$ (or ${\rm BR}_{\tau e}$) $\simeq0.86\%$ with \lumi~of collected data. The sensitivity degrades to 0.97\% when widening the $\slashed E_T$ distribution by 20\% (on top of the nominal detector simulation). We discuss additional potential sources for sensitivity detraction, such as the wrong identification of non-prompt leptons, when real experimental data is analyzed and all the systematic uncertainties are considered. Nevertheless, since the leading systematic uncertainty is expected to be governed by statistics, the sensitivity will improve as more data accumulates.

The existence of lepton flavor violating Higgs decays is an exciting possibility. Observation of such decay modes would call for physics beyond the SM related both to the electroweak symmetry breaking and to flavor physics. It might also shed light on other well-established observations that cannot be reconciled within the SM, such as neutrino oscillations and the baryon asymmetry of the Universe.

\vspace{0.5cm}
\begin{center}
{\bf Acknowledgements}
\end{center}

We thank Yevgeny Kats, Yossi Nir and Nadav Priel for useful discussions, and Dan Fishgold for his contribution.
This research was supported in part by the I-CORE Program of the planning and budgeting Committee and The Israel Science Foundation (grant NO 1937/12).


\end{document}